\documentclass[twocolumn,showpacs,preprintnumbers,amsmath,amssymb,aps,prl,superscriptaddress]{revtex4}

\usepackage{graphicx}
\usepackage{dcolumn}
\usepackage{bm}

\begin{document}

\title{Longitudinal Double-Spin Asymmetry and Cross Section for
Inclusive Jet Production in Polarized Proton Collisions at $\sqrt{s} = 200\,\mathrm{GeV}$}

\affiliation{Argonne National Laboratory, Argonne, Illinois 60439}
\affiliation{University of Birmingham, Birmingham, United Kingdom}
\affiliation{Brookhaven National Laboratory, Upton, New York 11973}
\affiliation{California Institute of Technology, Pasadena, California 91125}
\affiliation{University of California, Berkeley, California 94720}
\affiliation{University of California, Davis, California 95616}
\affiliation{University of California, Los Angeles, California 90095}
\affiliation{Carnegie Mellon University, Pittsburgh, Pennsylvania 15213}
\affiliation{University of Illinois, Chicago}
\affiliation{Creighton University, Omaha, Nebraska 68178}
\affiliation{Nuclear Physics Institute AS CR, 250 68 \v{R}e\v{z}/Prague, Czech Republic}
\affiliation{Laboratory for High Energy (JINR), Dubna, Russia}
\affiliation{Particle Physics Laboratory (JINR), Dubna, Russia}
\affiliation{University of Frankfurt, Frankfurt, Germany}
\affiliation{Institute of Physics, Bhubaneswar 751005, India}
\affiliation{Indian Institute of Technology, Mumbai, India}
\affiliation{Indiana University, Bloomington, Indiana 47408}
\affiliation{Institut de Recherches Subatomiques, Strasbourg, France}
\affiliation{University of Jammu, Jammu 180001, India}
\affiliation{Kent State University, Kent, Ohio 44242}
\affiliation{Institute of Modern Physics, Lanzhou, China}
\affiliation{Lawrence Berkeley National Laboratory, Berkeley, California 94720}
\affiliation{Massachusetts Institute of Technology, Cambridge, MA 02139-4307}
\affiliation{Max-Planck-Institut f\"ur Physik, Munich, Germany}
\affiliation{Michigan State University, East Lansing, Michigan 48824}
\affiliation{Moscow Engineering Physics Institute, Moscow Russia}
\affiliation{City College of New York, New York City, New York 10031}
\affiliation{NIKHEF and Utrecht University, Amsterdam, The Netherlands}
\affiliation{Ohio State University, Columbus, Ohio 43210}
\affiliation{Panjab University, Chandigarh 160014, India}
\affiliation{Pennsylvania State University, University Park, Pennsylvania 16802}
\affiliation{Institute of High Energy Physics, Protvino, Russia}
\affiliation{Purdue University, West Lafayette, Indiana 47907}
\affiliation{Pusan National University, Pusan, Republic of Korea}
\affiliation{University of Rajasthan, Jaipur 302004, India}
\affiliation{Rice University, Houston, Texas 77251}
\affiliation{Universidade de Sao Paulo, Sao Paulo, Brazil}
\affiliation{University of Science \& Technology of China, Hefei 230026, China}
\affiliation{Shanghai Institute of Applied Physics, Shanghai 201800, China}
\affiliation{SUBATECH, Nantes, France}
\affiliation{Texas A\&M University, College Station, Texas 77843}
\affiliation{University of Texas, Austin, Texas 78712}
\affiliation{Tsinghua University, Beijing 100084, China}
\affiliation{Valparaiso University, Valparaiso, Indiana 46383}
\affiliation{Variable Energy Cyclotron Centre, Kolkata 700064, India}
\affiliation{Warsaw University of Technology, Warsaw, Poland}
\affiliation{University of Washington, Seattle, Washington 98195}
\affiliation{Wayne State University, Detroit, Michigan 48201}
\affiliation{Institute of Particle Physics, CCNU (HZNU), Wuhan 430079, China}
\affiliation{Yale University, New Haven, Connecticut 06520}
\affiliation{University of Zagreb, Zagreb, HR-10002, Croatia}

\author{B.I.~Abelev}\affiliation{Yale University, New Haven, Connecticut 06520}
\author{M.M.~Aggarwal}\affiliation{Panjab University, Chandigarh 160014, India}
\author{Z.~Ahammed}\affiliation{Variable Energy Cyclotron Centre, Kolkata 700064, India}
\author{J.~Amonett}\affiliation{Kent State University, Kent, Ohio 44242}
\author{B.D.~Anderson}\affiliation{Kent State University, Kent, Ohio 44242}
\author{M.~Anderson}\affiliation{University of California, Davis, California 95616}
\author{D.~Arkhipkin}\affiliation{Particle Physics Laboratory (JINR), Dubna, Russia}
\author{G.S.~Averichev}\affiliation{Laboratory for High Energy (JINR), Dubna, Russia}
\author{Y.~Bai}\affiliation{NIKHEF and Utrecht University, Amsterdam, The Netherlands}
\author{J.~Balewski}\affiliation{Indiana University, Bloomington, Indiana 47408}
% ...

\collaboration{STAR Collaboration}\noaffiliation

\date{\today}

\begin{abstract}
We report a measurement of the longitudinal double-spin asymmetry
$A_{LL}$ and the differential cross section for inclusive midrapidity 
jet production in polarized proton collisions at $\sqrt{s}=200\,\mathrm{GeV}$.
The cross section data cover transverse momenta $5 < p_\mathrm{T} < 50\,\mathrm{GeV/c}$ 
and agree with next-to-leading order perturbative QCD  evaluations.
The $A_{LL}$ data cover $5 < p_\mathrm{T} < 17\,\mathrm{GeV/c}$ and disfavor at $98$\% C.L. 
maximal positive gluon polarization in the polarized nucleon.
\end{abstract}




\maketitle


Deep-inelastic scattering (DIS) experiments 
with polarized leptons and polarized nucleons have found that the spins of quarks 
and antiquarks account for only about 25\% of the nucleon spin~\cite{dis:0000}.
The gluon helicity distribution and orbital angular momenta are thus
essential to the understanding of the nucleon spin. 
Analyses of the scale dependence of the inclusive
nucleon spin structure function~\cite{scaling} and recent semi-inclusive DIS data~\cite{semi:0000}
have coarsely constrained the possible gluon spin contribution.
Complementary measurements with strongly interacting probes~\cite{E704:1991,Phenix:2005}
give sensitivity to gluons predominantly via quark-gluon and gluon-gluon scattering
contributions~\cite{Jager:2004}.

In this Letter we report the first measurement of $A_{LL}$ 
for inclusive jet production in polarized proton collisions,
\begin{equation}
A_{LL}= \frac{{\sigma^{++}-\sigma^{+-}}}{{\sigma^{++}+\sigma^{+-}}},
\label{all:eq}
\end{equation}
where $\sigma^{++}$ and $\sigma^{+-}$ are the inclusive jet cross sections when the two 
colliding proton beams have equal and opposite helicities, respectively.
In addition we report the inclusive jet differential cross section.

In pQCD the (un-)polarized jet cross section involves a convolution of (un-)polarized 
quark and gluon distribution functions and the (un-)polarized hard partonic scattering cross 
section~\cite{Jager:2004,RHIC:Spin}.  
We compare next-to-leading order (NLO) pQCD calculations with the measured cross section
to test their applicability and to support their use in constraining the polarized gluon 
distribution through measurement of $A_{LL}$. Our data on $A_{LL}$
are sensitive to gluon polarization for momentum fractions $0.03<x<0.3$.

The data were collected at the Brookhaven Relativistic Heavy Ion Collider (RHIC) with 
the Solenoidal Tracker at RHIC (STAR)~\cite{nim:2003} in the years 2003 and 2004
using proton beams of 100 GeV energy.
Typical luminosities were $2$--$5\times10^{30}\,\mathrm{cm}^{-2}\mathrm{s}^{-1}$.
Spin rotator magnets upstream and downstream of the STAR interaction region (IR) rotated the proton 
beam spins from and to the stable vertical direction in RHIC to provide collisions with 
longitudinal polarizations~\cite{nim:2003}.
The helicities alternated for successive bunches of one beam and for successive pairs of bunches of the other beam.
Thus STAR recorded collisions with all beam helicity combinations in rapid succession.

The polarization of each beam was measured for each beam fill 
with RHIC Coulomb-Nuclear Interference (CNI) proton-carbon polarimeters~\cite{CNI:2003},
which were calibrated \emph{in situ} using a polarized atomic hydrogen gas-jet target~\cite{JET:2006}.
Proton beam polarizations were $30$\%--$45$\%.
Nonlongitudinal beam polarization components at the STAR IR were measured 
continuously with local polarimeters~\cite{BBC:2004} and were no larger than $9$\% (absolute).

The STAR detector subsystems~\cite{nim:2003} of principal interest here are the time projection chamber (TPC), 
the barrel electromagnetic calorimeter (BEMC), and the beam-beam counters (BBC).
The TPC tracks charged particles in a $0.5$ T solenoid magnetic field for all azimuthal angles ($\phi$) 
and pseudorapidities $|\eta| \lesssim1.3$.
The BEMC is a lead-scintillator sampling calorimeter 
that limited the acceptance in 2003 and 2004,
covering all $\phi$ and $0 < \eta < 1$ with respect to the TPC center. 
The BBCs are composed of segmented scintillator annuli that span $3.3 < |\eta| < 5.0$ and measure the proton 
beam luminosity and transverse polarization components.

Proton collision events were identified by coincident signals from at least one BBC segment 
on either side of the IR. The cross section for the BBC coincidence requirement is 
$26.1\pm2.0\,\mathrm{mb}$, which is $87$\% of the non-singly diffractive pp cross section~\cite{Gans}.
The jet data were collected with a highly prescaled minimum bias (MB) trigger, 
requiring a proton collision event, and a high tower (HT) calorimetric trigger condition.
It required, in addition, a signal
from at least one BEMC tower of size $\Delta \eta \times \Delta \phi = 0.05 \times 0.05$ above 
a transverse energy ($E_\mathrm{T}$) threshold of $2.2\,\mathrm{GeV}$ in 2003 
($2.2$--$3.4\,\mathrm{GeV}$ at $\eta=0$--$1$ in 2004).
In total $2.1\times10^{6}$ MB and $3.0\times10^{6}$ HT events were analyzed.
The integrated luminosity $\int \mathcal{L} {\rm{dt}}$ amounts to $0.18\,(0.12)\,\mathrm{pb}^{-1}$ 
for the analyzed 2003 (2004) data.

Jets were reconstructed using a midpoint-cone algorithm~\cite{CDF:2000} that clusters 
reconstructed TPC tracks and BEMC energy deposits within a cone 
in $\eta$ and $\phi$ starting from energy seeds of at least $0.5\,\mathrm{GeV}$.
A cone radius $r_{\rm{cone}}=0.4$ was chosen because of the limited BEMC $\eta$ acceptance.
Particle tracks with $p_\mathrm{T}>0.2\,\mathrm{GeV/c}$ were considered if they originated 
from the primary interaction vertex, which 
was required to be on the beam axis and within 
$60\,\mathrm{cm}$ from the TPC center to ensure
uniform tracking efficiency. Calorimeter towers were considered 
if their $E_\mathrm{T}$ exceeded $0.2\,\mathrm{GeV}$ after correction for charged hadron 
contributions determined from TPC tracking.  A charged pion (photon) mass was 
assumed for tracks (towers) in relating energy and momentum. Jets were required to have 
a reconstructed jet $p_\mathrm{T} > 5\,\mathrm{GeV/c}$ and, as a tradeoff between acceptance 
and effects from acceptance edges, a reconstructed jet  
axis intersecting the BEMC at nominal $\eta$ between $0.2$ and $0.8$. 
A minimum TPC contribution to the jet energy, $E_{\rm{TPC}}/E_{\rm{tot}} > 0.2~(0.1)$ in 2003 (2004), 
was used to suppress apparent jets from beam background.
The jet $p_{{\rm{T}}}$ resolution was determined to be $\sim$25\% from the momentum 
balance of dijet events and from simulation, and motivated the choice of binning.

\begin{figure}
  \includegraphics[width=8.5cm,clip]{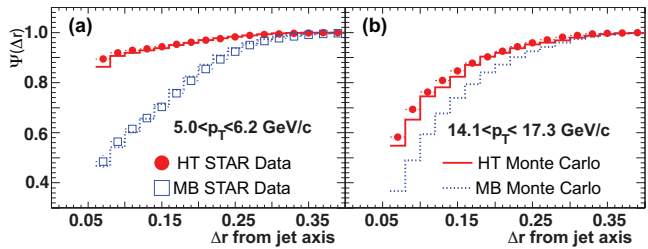}
  \caption{Jet profile $\Psi(\Delta r,r_{\rm{cone}},p_\mathrm{T})$ versus inner cone size 
    $\Delta r$ at $r_{\rm{cone}}= 0.4$ for MB (open squares) and HT (filled circles) 
    data compared with STAR Monte Carlo simulation in two jet $p_\mathrm{T}$ bins 
    (a) $5.0 < p_\mathrm{T} < 6.2$ and (b) $14.1 < p_\mathrm{T} < 17.3\,\mathrm{GeV/c}$.
    In (b) the MB jet yield was too small to measure.} 
\label{fig:profile}
\end{figure}

Figure~\ref{fig:profile} compares the measured and simulated jet profile
$\Psi(\Delta r,r_{\rm{cone}},p_\mathrm{T})$,
defined as the average fraction of jet $E_\mathrm{T}$ inside a coaxial inner cone
of radius $\Delta r < r_\mathrm{cone}$, for the MB and HT data separately.
The reconstruction software imposed the same trigger requirements as in the data.
More than $95$\% of the total jet energy is expected to be contained within $r_{\rm{cone}}=0.4$.
The HT trigger, providing increased selectivity for jets, causes a $p_{{\rm{T}}}$
dependent bias toward jets with hard fragments that produce an electromagnetic shower.
The $\Psi(\Delta r,r_{\rm{cone}},p_\mathrm{T})$ distributions are well reproduced by
\textsc{pythia}-based (v 6.205~\cite{pythia} `CDF TuneA' settings~\cite{CDF:TuneA})
Monte Carlo simulations passed through \textsc{geant}-based~\cite{geant:321} STAR detector
simulation. The simulations are used in determining the cross section and to assess effects 
of the trigger bias on $A_{LL}$. In the cross section analysis of HT data an $E_\mathrm{T}$ 
threshold of $3.5\,\mathrm{GeV}$ was imposed on the BEMC trigger tower to ensure 
a uniform trigger efficiency.

The differential inclusive cross sections were determined separately for the MB and HT data according to
\begin{equation}
\frac{1}{2\pi} 
\frac{{\rm{d^2}}\sigma }{ {\rm{d}}\eta  {\rm{d}} p_{{\rm{T}}} } =
\frac{1}{2\pi} 
\frac{N_{\rm{jets}}}{\Delta \eta \Delta p_{{\rm{T}}} } 
\frac{1}{\int \mathcal{L} {\rm{dt}}} 
\frac{1}{c(p_{{\rm{T}}}) },
\end{equation}
where $N_{\rm{jets}}$ denotes the number of jets 
observed within a pseudorapidity interval $\Delta \eta$ and a transverse momentum interval 
$\Delta p_{{\rm{T}}}$ at a mean jet $p_{{\rm{T}}}$.
The correction factors $c(p_{{\rm{T}}})$ were determined from simulation, and are defined 
as the ratio of the number of jets reconstructed within a given $p_{{\rm{T}}}$ interval in the simulated 
data to those generated in the \textsc{pythia} final-state particle record.
They change monotonically for HT events from $0.02$ at $p_\mathrm{T}=8.3\,\mathrm{GeV/c}$ 
to $0.79$ at $p_\mathrm{T}=43\,\mathrm{GeV/c}$, whereas they are a constant $0.69$ 
for MB events with $p_\mathrm{T}<12.6\,\mathrm{GeV/c}$. 
Consistent values were obtained with the \textsc{herwig}~\cite{herwig}~generator. 
Typically $35$\%-$40$\% of the jets generated in a given $p_\mathrm{T}$ interval were reconstructed
in the same interval.  Reconstructed $p_\mathrm{T}$ was found to be
on average $\sim$20\% larger than generated $p_\mathrm{T}$ in each reconstructed 
$p_\mathrm{T}$ interval, and the difference is taken into account via~$c(p_\mathrm{T})$.

The MB differential cross sections extracted from $1.4\times10^3$ jets collected in 2003 
and $1.1\times10^3$ in 2004 are in good agreement ($\chi^2/\mathrm{ndf} = 0.8$).
A $20$\% systematic offset 
for all  $p_\mathrm{T}$ was found between 
the HT differential cross sections extracted from $43\times10^3$ and $42\times10^3$ jets 
collected in 2003 and 2004. 
We ascribe this difference to $5$\% uncertainty (included in the systematic errors below) 
in the year-to-year absolute scale of the BEMC calibration, 
which was changed by a factor of $\sim2$ between the two years,
and to uncertainty
in the modeling of temporary BEMC hardware malfunctions.
The calibration used $20\times10^6$ d+Au collision events in 2003 and $50\times10^6$ Au+Au events in 2004.
The absolute energy scale was set by matching BEMC energy to TPC track momentum for well-contained 
showers from $1.5< p <8\,\mathrm{GeV/c}$ electrons identified in the TPC.
Uncertainties arise in the electron selection, from residual hadronic contamination, 
and from the limited d+Au statistics.

\begin{figure}
  \includegraphics[width=7.5cm,clip]{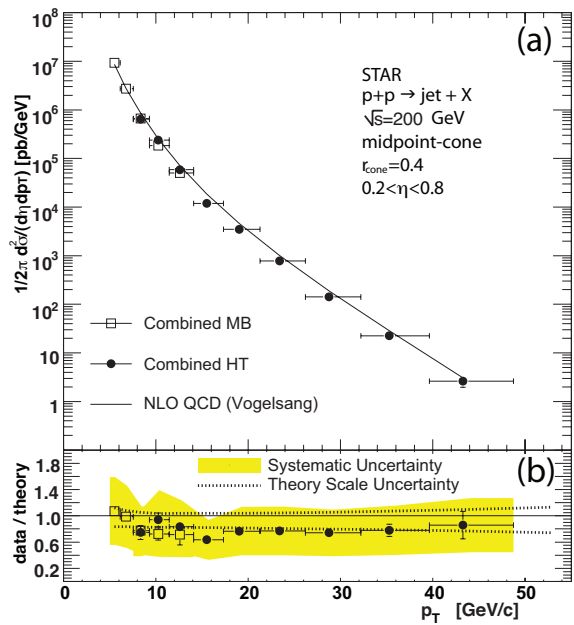}
  \caption{\label{fig:cs}
  (a) Inclusive differential cross section for $p+p \rightarrow {\rm{jet}} +X$ 
at $\sqrt{s} = 200\,\mathrm{GeV}$ versus jet $p_\mathrm{T}$ for a jet cone radius 
of $0.4$. The symbols show MB (open squares) and HT (filled circles) data 
from the years 2003 and 2004 combined. The horizontal bars indicate the ranges 
of the $p_\mathrm{T}$ intervals. The curve shows a NLO calculation\cite{Jager:2004}.
(b) Comparison of theory and data. The band indicates the experimental systematic uncertainty.
The upper (lower) dashed line indicates the relative change of the NLO calculation 
when it is evaluated  at $\mu = p_\mathrm{T}/2$ ($\mu = 2p_\mathrm{T}$).}
\end{figure}

Figure~\ref{fig:cs}(a) shows the arithmetic average of the 2003 and 2004 MB and HT cross sections
versus jet $p_\mathrm{T}$.
The MB and HT data are in good agreement for 
overlapping jet $p_\mathrm{T}$ ($\chi^2/\mathrm{ndf} = 1.0$), despite the very different $c(p_\mathrm{T})$.
The curve shows the NLO pQCD cross section of Ref.~\cite{Jager:2004} 
evaluated at equal factorization and renormalization scales, 
$\mu\equiv\mu_F=\mu_R=p_\mathrm{T}$, using the CTEQ6M 
parton distributions~\cite{CTEQ6:2004}.
Figure~\ref{fig:cs}(b) compares data and theory, showing
satisfactory agreement over $7$ orders of magnitude.
The theoretical cross section changes by less than $23$\% if
$\mu$ is varied by a factor of two and increases by $1$\% ($13$\%) at $p_\mathrm{T}$ of
$10$ ($40$)$\,\mathrm{GeV/c}$ if the CTEQ6.1M distributions are used.
The experimental systematic uncertainty amounts to $8\%$ in the normalization with the
BBC and $48$\% in the measured yield, consisting of $5$\% due 
to residual beam background, $13$\% on $c(p_\mathrm{T})$, and $46$\% from 
a $9$\% uncertainty on the jet energy scale.
The BEMC calibration and undetected neutral particles dominate in the latter.
No corrections were made for the nonperturbative redistribution of energy into and out of the jet 
by the underlying event and out-of-cone hadronization. 
We estimate that such corrections would increase the measured differential cross section by $\sim$25\% 
for $p_\mathrm{T} > 10\,\mathrm{GeV/c}$. 


The asymmetry $A_{LL}$ was extracted for $5 < p_\mathrm{T} < 17\,\mathrm{GeV/c}$ from a HT  
data sample of about $110\times10^3$ jets in 2003 and $210\times10^3$ in 2004.
The sample size is larger than in the cross section analysis, since no BEMC energy threshold was required.
The jet yields $N$ were sorted by equal $(++)$  and opposite $(+-)$ beam helicity configurations.
The asymmetry was extracted as:
\begin{equation}
\hspace{-0.9cm}A_{LL}= \frac{\sum (P_1 P_2) (N^{++} - R N^{+-}) }{\sum (P_1 P_2)^2 (N^{++} + R N^{+-}) }
\label{eq:all}
\end{equation}
\noindent
where $P_{1,2}$ are the measured proton beam polarizations, $R\simeq1.1$ is 
the ratio of measured luminosities for equal and opposite proton beam helicities,
and parity violating differences $\lesssim \mathcal{O}(10^{-4})$ in the cross
sections for different beam helicities are not considered. 
The sums are performed over runs typically lasting 20 minutes.

\begin{figure}
  \includegraphics[width=8.0cm,clip]{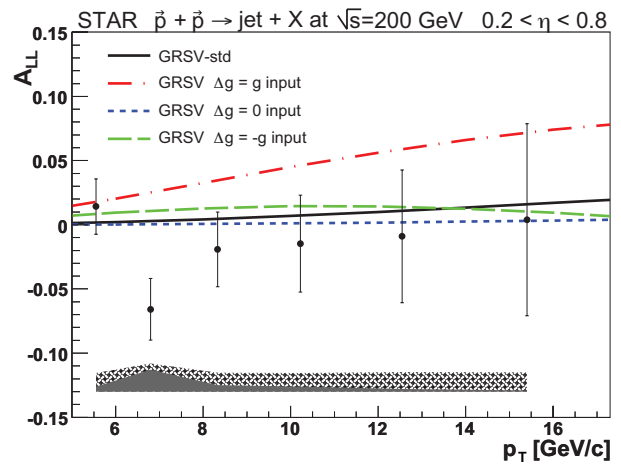}
   \caption{\label{fig:all} The longitudinal double-spin asymmetry $A_{LL}$ in
   $\vec{p}+\vec{p} \rightarrow {\rm{jet}} +X$ at $\sqrt{s} = 200\,\mathrm{GeV}$ 
   versus jet $p_\mathrm{T}$. The uncertainties on the data points are statistical.
   The gray band indicates the systematic uncertainty from the beam polarization 
   measurement, and the hatched band the total systematic uncertainty.
   The curves show predictions based on deep-inelastic scattering 
   parametrizations of gluon polarization~\cite{Jager:2004,CTEQ6:2004}.
   }
\end{figure}

The results for $A_{LL}$ from 2003 and 2004 data are in good  agreement 
($\chi^2/\mathrm{ndf} = 0.3$).
Figure~\ref{fig:all} shows the combined $A_{LL}$ versus jet $p_\mathrm{T}$, 
together with the statistical (bars) and systematic (bands) uncertainties.

A $25$\% combined scale uncertainty arises from the CNI beam polarization  
measurement ($22$\% in 2003 and an uncorrelated $16$\% in 2004) and  
from the CNI absolute calibration ($18$\% common to both years).

The uncertainty in $R$ was estimated to be $0.003$ using narrow and  
wide timing requirements for the BBC coincidence. 
It takes into account differences in sampling of the longitudinal vertex
distribution in the jet analysis and in the relative luminosity  
measurement, and corresponds to $0.009$ uncertainty in $A_{LL}$.  
An  independent measurement with the zero degree calorimeters (ZDC)~\cite{nim:2003} 
gave consistent results to within statistical uncertainties.
No double helicity asymmetry of the BBC measurement relative to the  
ZDC measurement was observed.

Residual nonlongitudinal proton beam polarization at the STAR IR could contaminate 
the $A_{LL}$ measurement through an azimuthally uniform two-spin asymmetry~\cite{ASigma:1998}.
A limit of $0.010$ on such contamination
was set from local polarimetry data and from two-spin asymmetry measurements 
with vertically polarized beams.

Beam background occasionally caused BEMC signals not associated with  
collisions at the IR.
Its effect on the jet yields was reduced with the aforementioned  
selection on $E_{\rm{TPC}}/E_{\rm{tot}}$.
Residual yields were estimated to be no larger than $8$\% ($5$\%) in  
the $2003$ ($2004$) data from the variation of jet spectra with beam-background 
conditions monitored with the BBCs when filled and empty  
beam bunches crossed at the IR.
These, combined with asymmetry estimates from beam background  
dominated samples, resulted in $0.003$ uncertainty in $A_{LL}$.

The bias toward hard fragmentation processes caused by the HT trigger requirement 
was simulated, as were possible biases introduced by jet reconstruction and jet $p_\mathrm{T}$ resolution.
The resulting $p_\mathrm{T}$ dependent shifts in $A_{LL}$ were estimated 
with the polarized parton distributions of Ref.~\cite{GRSV:2000}.
Their total is estimated to be less than $0.009$.

Analyses with randomized proton beam helicity configurations and 
other cross checks including parity violating single-spin asymmetries 
showed the expected statistical behavior, thus indicating no evidence 
for beam bunch to bunch or fill to fill systematics in $A_{LL}.$


The curves in Figure~\ref{fig:all} show theoretical evaluations~\cite{Jager:2004,CTEQ6:2004} 
at $\mu=p_\mathrm{T}$ for the commonly used polarized parton distributions 
of  Ref.~\cite{GRSV:2000}. They shift by less than $0.003$ ($0.017$) 
at $p_\mathrm{T} = 5.6$ ($15.7$) $\mathrm{GeV/c}$ if $\mu$ is varied by a factor of $2$.
The polarized parton distributions are based on a best fit to polarized inclusive DIS data, 
the so-called GRSV-standard 
gluon polarization distribution, and on assumptions of (i) a vanishing gluon polarization  
$\Delta g(x,Q^2_0) = 0$, and (ii) maximally positive or negative gluon polarization, 
$\Delta g(x,Q^2_0) = \pm g(x,Q^2_0)$, at the initial scale $Q^2_0=0.4\,\mathrm{GeV}^2/\mathrm{c}^2$ 
in the analysis~\cite{GRSV:2000}. 
Alternative parametrizations~\cite{AAC:2004} are within the range spanned by (ii).
Our data fall below the
$\Delta g(x,Q^2_0)=g(x,Q^2_0)$ evaluation ($\chi^2/\mathrm{ndf}\simeq3$) and are consistent 
with the other evaluations, in qualitative agreement with Refs.~\cite{E704:1991,semi:0000,Phenix:2005}.
The results thus disfavor large and positive gluon polarization, 
proposed~\cite{altarelli:1988} originally to explain the small quark spin contribution to the proton spin.

In summary, we report the first measurement of the  longitudinal double-spin asymmetry $A_{LL}$ 
for inclusive jets with transverse jet momenta of $5 < p_\mathrm{T} < 17\,\mathrm{GeV/c}$ produced at 
midrapidity in polarized proton collisions at $\sqrt{s}=200$ GeV.
The jet cross section was determined for $5 < p_\mathrm{T} < 50\,\mathrm{GeV/c}$ 
and is described by NLO pQCD evaluations over $7$ orders of magnitude.
The asymmetries $A_{LL}$ are consistent with NLO pQCD calculations utilizing polarized quark and
gluon distributions from inclusive DIS analyses,
and disfavor at $98$\% C.L. large positive values of gluon polarization in the polarized nucleon.

We thank the RHIC Operations Group and RCF at BNL, and the
NERSC Center at LBNL for their support. This work was supported
in part by the Offices of NP and HEP within the U.S. DOE Office 
of Science; the U.S. NSF; the BMBF of Germany; CNRS/IN2P3, RA, RPL, and
EMN of France; EPSRC of the United Kingdom; FAPESP of Brazil;
the Russian Ministry of Science and Technology; the Ministry of
Education and the NNSFC of China; IRP and GA of the Czech Republic,
FOM of the Netherlands, DAE, DST, and CSIR of the Government
of India; Swiss NSF; the Polish State Committee for Scientific 
Research; SRDA of Slovakia, and the Korea Sci. \& Eng. Foundation.

\end{document}